\documentclass[authoryear,12pt,letter,3p]{jowarticle}
\usepackage{graphicx}
\usepackage{amsmath}
\usepackage{amssymb}
\usepackage{natbib}
\usepackage{setspace}
\usepackage[all]{xy}
\usepackage{enumitem}
\usepackage{titlesec}
\makeatletter
\renewcommand\section{\@startsection{section}{1}{\z@}{-3.25ex plus -1ex minus -.2ex}{1.5ex plus .2ex}{\normalsize\bf}}
\renewcommand\subsection{\@startsection{subsection}{2}{\z@}{-3.25ex plus -1ex minus -.2ex}{1.5ex plus .2ex}{\normalsize\bf}}
\renewcommand\subsubsection{\@startsection{subsubsection}{3}{\z@}{-3.25ex plus -1ex minus -.2ex}{1.5ex plus .2ex}{\normalsize\bf}}
\makeatother

\begin{document}
\begin{frontmatter}
\title{Classical Spacetime Structure}

\author{James Owen Weatherall}\ead{weatherj@uci.edu}
\address{Department of Logic and Philosophy of Science\\ University of California, Irvine}
\begin{abstract}I discuss several issues related to ``classical'' spacetime structure.  I review Galilean, Newtonian, and Leibnizian spacetimes, and briefly describe more recent developments.  The target audience is undergraduates and early graduate students in philosophy; the presentation avoids mathematical formalism as much as possible.\end{abstract}
\end{frontmatter}
\doublespacing
\section{Introduction}\label{introduction}

One often associates \emph{spacetime}---a four-dimensional geometrical structure representing both space and time---with relativity theory, developed by Einstein and others in the early part of the twentieth century.\footnote{The idea of ``spacetime'' was actually introduced by \citet{Minkowski}, several years after Einstein first introduced relativity theory.} But soon after relativity theory appeared, several authors, such as Hermann \citet{Weyl}, \'{E}lie \citet{Cartan1, Cartan2}, and Kurt \citet{Friedrichs}, began to study how the spatio-temporal structure presupposed by classical, i.e., Newtonian, physics could be re-cast using the methods of four-dimensional geometry developed in the context of relativity.

These reformulations of classical physics were initially of interest for the insight they offered into the structure of relativity theory.\footnote{And indeed, they remain of interest for this reason: see, for instance, \citet{Friedman}, \citet{MalamentLimit1,MalamentLimit2,MalamentGR}, \citet{EarmanWEST}, \citet{FletcherLimit}, \citet{BarrettSTS}, \citet{WeatherallExplanation,WeatherallSGP,WeatherallSingularity,WeatherallPuzzleBall,WeatherallConservation}, and \citet{Weatherall+Dewar} for philosophical discussions of the relationship between relativity theory and Newtonian physics that make heavy use of this formalism.}  But this changed in 1967, when Howard Stein \nocite{SteinNST} proposed that the notion of a ``classical'' spacetime could provide important insight into Newtonian physics itself---including on issues that preoccupied Newton and his contemporaries regarding the status of ``absolute'' space and motion.  On Stein's reconstruction, Newton's oft-decried ``absolutism'' about space amounted to the claim that the laws governing the motion of bodies presupposed that there are (unobservable) facts about which bodies are at rest.  This idea can be reconstructed as a claim that space and time together have a certain geometrical structure, now known as \emph{Newtonian spacetime}.  Leibniz's response, meanwhile, that there was no discernible difference between whether all of the bodies in the world were at rest or in uniform rectilinear motion, can then be taken as an argument that Newton's laws assume more structure than can be supported on metaphysical grounds; Leibniz, it seems, believed space and time had the structure of \emph{Leibnizian spacetime}, which, in a certain precise sense, is \emph{less structure} than Newtonian spacetime.\footnote{For more on the notion of ``structure'' being used here, see \citet{BarrettSTS} and \citet{WeatherallCategories}.  \label{EarmanStein} Newtonian spacetime was introduced in Stein's original 1967 paper; Leibnizian spacetime was introduced by Stein somewhat later \citep{SteinPrehistory}, and then developed by \citet{Earman1977, Earman1979, Earman1986, EarmanWEST, Earman1989}.  Note, however, that there are subtle differences in Earman and Stein's understanding of what Leibnizian spacetime captures \citep{WeatherallEarmanStein}, and that Earman, in particular, believed that Leibnizian spacetime includes an implicit commitment to some form of ``subtantivalism'' (see footnote \ref{substantivalism}) that Leibniz would have denied; he suggested that one should move to an algebraic framework to better reflect Leibniz's views.  For replies to this proposal, see \citet{Rynasiewicz1992} and \citet{Rosenstock+etal}.}

Perhaps the most striking and influential aspect of Stein's paper was his argument that in fact the spatio-temporal structure presupposed by Newton's laws of motion was somewhat less than Newton imagined---though somewhat more than Leibniz would have accepted.  This intermediate structure has come to be known as \emph{Galilean spacetime}.\footnote{What we now call Galilean spacetime was first discussed by \citet{Weyl}.  It is not clear whether any of Newton's contemporaries were in a position to recognize this intermediate structure, though \citet{SteinNST,SteinPrehistory} and others have suggested that Christian Huygens came closest. See \citet{Stan} for a recent discussion of Huygens' views on rotation, which is the context in which he most closely approached the idea of Galilean spacetime.}  Remarkably, Galilean spacetime provides the resources needed to both avoid Leibniz's famous shift argument,\footnote{What I call the shift argument concerns setting the whole world in motion at a constant velocity; it is sometimes known as the ``kinematic'' shift argument, to distinguish it from the ``static'' shift argument, wherein one considers shifting the entire universe by some fixed amount \citep{PooleyOxford}; see also footnote \ref{substantivalism}.} which exposes the unobservability of absolute uniform rectilinear motion, and also to accept Newton's famous bucket argument, which Newton and many others have taken to show that at least \emph{some} absolute motions are empirically detectable.

The remainder of this chapter will proceed as follows.\footnote{My presentation will not be particularly historically sensitive; nor will it be technical.  For more historical detail, but with the same basic perspective, see \citet{WeatherallVoid} and (especially) references therein; for technical details, see \citet{WeatherallSTG}.  For a treatment at a level comparable to the one attempted here, which also develops ideas from relativity theory, see \citet{GerochAB}; a more philosophical perspective is offered by \citet{Maudlin}.}  I will begin by making some remarks about ``space'' in classical physics.  I will then introduce the notion of Galilean spacetime and discuss several senses in which it is the spacetime structure presupposed by Newtonian physics.  Next I will discuss Newtonian spacetime and Leibnizian spacetime.  I will conclude by briefly discussing some other ideas concerning classical spacetime structure, including the recent proposal by \citet{Saunders} that one should take a structure strictly intermediate between Galilean and Leibnizian spacetimes to be what is really presupposed by Newtonian physics.

\section{Space}\label{space}

Before discussing classical space\emph{time}, we first consider the geometry of space---something that all of the structures I will discuss below agree on.  \emph{Space}, in what follows, will be understood as a collection of (infinitesimally) small \emph{places}, i.e., locations where small bodies may be situated.  We will call these locations the \emph{points} of space.\footnote{\label{substantivalism} There is a subtle metaphysical issue here, concerning whether we understand these ``points'' or ``places'' to themselves be physical objects, existing (metaphysically) prior to and independently of bodies (\emph{substantivalism}); or if instead they characterize something about relations between bodies (\emph{relationism}).  Nothing I say here should be understood to be taking one or the other of these positions for granted.  Perhaps more importantly, the classical spacetime structures I describe here are supposed to provide insight on a different issue, concerning the character of \emph{motion}, rather than, in the first instance anyway, the metaphysics of space or time.  See \citet{WeatherallHoleArg,WeatherallEarmanStein} for more on this perspective; for further discussion and other perspectives, see \citet{EarmanWEST}, \citet{BelotGM}, and \citet{PooleyOxford}.}   (Since most physical objects are not vanishingly small, their locations cannot be represented by single points of space; in general, bodies occupy regions of space, i.e., collections of contiguous points.)   This collection of points is understood to be structured, in the sense that there are various further relations defined on them.  It is these relations that one aims to characterize when one speaks of ``spatial structure''---or, \emph{mutatis mutandis}, ``spatio-temporal structure''.

\begin{figure}
\begin{centering}
\includegraphics[width=\columnwidth]{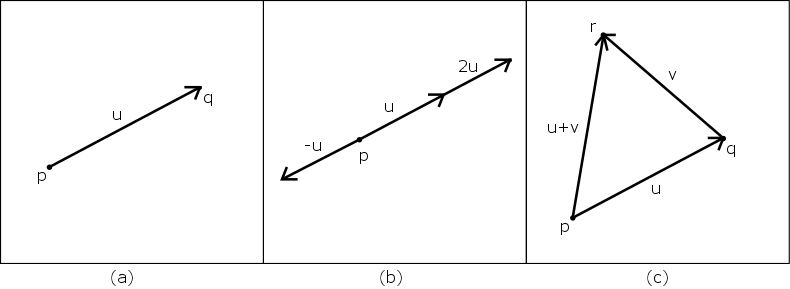}
\end{centering}
\caption{\label{vectors} (a) An arrow $u$ relating two points $p$ and $q$ of space. (b) ``Scaling'' an arrow $u$ at a point $p$ by various amounts, including flipping its direction.  (c) Adding arrows $u$ and $v$: here $u+v$ is the arrow relating $p$ are $r$, where $r$ is the point determined by $q$ and $v$, and $q$ is the arrow determined by $p$ and $u$.}
\end{figure}

For example, space is three dimensional.  We make this idea precise by introducing the notion of an \emph{arrow} between pairs of points of space.\footnote{These arrows will make space a (three dimensional) \emph{affine space}.  See \citet{MalamentGST}, or \citet{WeatherallSTG} for details.} (See Fig. \ref{vectors}(a).)  First, pick any two points of space, $p$ and $q$.  We suppose one can always draw a (unique) arrow whose tail begins at $p$ and whose head lands at $q$, and likewise, one can always draw an arrow whose tail begins at $q$ and whose head ends at $p$.  Conversely, we suppose that any arrow whose tail begins at $p$ ends somewhere in space, i.e., at some point.  This means that we can think of any point and an arrow originating at that point as uniquely picking out another point of space; and we can think of any ordered pair of points as uniquely determining an arrow.

We will call arrows that originate at a point $p$ \emph{arrows at $p$}.  We suppose that given an arrow at $p$, one can identify the ``same'' arrow---i.e., an arrow with the same length and orientation---at any other point.  Thus it makes sense to speak of an arrow without mentioning the point at which it is based.

Now pick any point of space, $p$.  We define two basic ways of manipulating the arrows at $p$.  For one, we assume we can always ``scale'' or any arrow.  (See Fig. \ref{vectors}(b).)  That is, given any arrow $u$ (say), one can uniquely identify an arrow that points in the same direction as $u$, but which is twice as long (say).  We will call this arrow $2u$.  Likewise, one can identify an arrow that points in precisely the opposite direction as $u$; we will call this $-u$, short for $(-1)u$. Given any arrow $u$ at $p$, $0u$ is the (unique) arrow from $p$ to itself.

The second operation one can perform on arrows at a point is to add them to one another.  Consider two arrows $u$ and $v$ at $p$.  We define the arrow $u+v$ to be the unique arrow from $p$ to the point $r$, where $r$ is defined as follows: first, consider the point $q$ determined by $p$ and $u$; then, take $v$, and let $r$ be the point determined by $q$ and $v$. (See Fig. \ref{vectors}(c).) In other words, we understand $u+v$ to be the arrow taking us from $p$ to the point we would reach by first following $u$ and then following $v$.  We assume that for all $u$ and $v$, $u+v=v+u$---i.e., following $v$ and then $u$ leads to the same place as following $u$ and then $v$.\footnote{The arrows at each point are to be understood as forming a vector space; and there is a canonical isomorphism between the vector spaces at each point.  Vector spaces are required to satisfy some additional conditions that I do not mention, but these are all met by the mental model meant to be invoked by this description.  See the texts already cited for details.}

I can now say what it means for space to be three dimensional.  Space is three dimensional just in case at any point $p$, there exist three arrows, $x$, $y$, and $z$, which are such that (1) none of these three can be constructed by any process of scaling or adding the other two; and (2) one can construct any arrow $v$ at $p$ just by scaling and/or adding together $x$, $y$, and $z$.  The important thing, here, is that for ordinary space, any collection of arrows with these two properties always has exactly the same number of elements: namely, three.

There is a bit more structure that we will take space to have.  First, given any two points $p$ and $q$, we assume we can say how far apart they are: that is, we have a notion of \emph{spatial distance}.\footnote{This notion of spatial distance consists in a Euclidean metric on the vector space of arrows between points.}  In other words, we assign a \emph{length}---a non-negative number, which is zero only for the arrow taking a point to itself---to the arrow $v$ between $p$ and $q$, which we will write $||v||$.  Likewise, given two arrows $u$ and $v$, we can assign an angle---a number between $0$ and $2\pi$---to them, where an angle of $0$ (or $2\pi$) means the arrows point in the same direction; an angle of $\pi$ means they point in opposite directions; and an angle of $\pi/2$ or $3\pi/2$ means they are \emph{orthogonal}, i.e., the angle between $u$ and $v$ is precisely the same as the angle between $u$ and $-v$ (or, equivalently, between $-u$ and $v$).  Finally, we assume that this notion of length satisfies the following two conditions: (a) for any real number $a$ and any arrow $u$, $||au|| = |a|\cdot||u||$, where $|a|$ is the absolute value of $a$; and (b) if $u$ and $v$ are orthogonal, then $||u+v||^2 = ||u||^2 + ||v||^2$.

\section{Galilean Spacetime}

We now turn to our first spacetime structure: \emph{Galilean spacetime}.  To characterize Galilean spacetime, we begin much as in our discussion of space: Galilean spacetime consists in a collection of points, now understood as locations not only in space, but also in time.  Rather than ``places'', these locations are \emph{events}, in the sense that they represent occurrences in a small region of space for an instant of time.  They are not necessarily \emph{significant} events: an event might be a speck of dust existing at a moment, or even the occurrence of nothing at all.

Just as ordinary extended objects are not located at single points of space, neither are objects that are extended in space---say, a rope---nor objects that persist through time---say, a particle or a person---represented by single points in spacetime.  Instead, these are represented by various sorts of curves and surfaces, as described below.

This collection of events is once again structured.  In particular, Galilean spacetime is a four-dimensional ``space'' of events, in much the same way that space is three dimensional.  That is, we suppose that any pair of spacetime points $p$ and $q$ are uniquely related by an arrow; and that given any point $p$, and any arrow $v$, there is a unique point $q$ such that $v$ is the arrow from $p$ to $q$.  Now, though, we suppose that at any point $p$ of Galilean spacetime, one can find four arrows---$t$, $x$, $y$, and $z$---with the properties that (1) none of these four can be constructed by any process of scaling or adding the other three; and (2) one can construct any other arrow by scaling and/or adding together $t$, $x$, $y$, and $z$.\footnote{In other words, Galilean spacetime consists in a four-dimensional affine space of events, with further structure to be described.}

Now consider any two points $p$ and $q$ in Galilean spacetime.  There are several additional relationships that hold between them.  One such notion is \emph{temporal distance}, $t$, which assigns a real number to any ordered pair of events---or, equivalently, a \emph{temporal length} to any arrow relating events.  This number represents the duration between those two events.  Temporal distance has the following properties.  First, if $u$ and $v$ are arrows, then $t(u+v) = t(u) + t(v)$, i.e., if $p$ and $q$ are related by $u$, and $q$ and $r$ are related by $v$, then the temporal distance between $p$ and $r$ is the sum of the temporal distance between $p$ and $q$ and the temporal distance between $q$ and $r$.  Likewise, given any real number $a$ and any vector $u$, the temporal distance satisfies $t(au)=at(u)$, where on the left hand side of this equation we are applying the scaling operation, and on the right hand side we are just multiplying real numbers.\footnote{That is, temporal duration is a linear functional acting on the vector space associated with Galilean spacetime.}  Finally, we can always find a (non-unique) collection of three arrows, $x$, $y$, and $z$, with the properties that (1) none of these three can be constructed by any process of scaling or adding the other two; and (2) the temporal distance assigned to each of these arrows, and thus all other arrows that can be constructed by scaling and adding them, is zero.  

If the temporal distance from $p$ and $q$ is positive, we say that $q$ is in the \emph{future} of $p$; if it is negative, we say it is in the \emph{past} of $p$; if it is zero, then $p$ and $q$ are \emph{simultaneous}. Now let $p$ be a point, and let $x$, $y$, and $z$ be three arrows satisfying (1) and (2) in the previous paragraph.  Then any point $q$ related to $p$ by an arrow that can be constructed by scaling or adding $x$, $y$, and $z$ will be simultaneous with $p$, and all events simultaneous with $p$ can be found in this way.  Thus, the events simultaneous with $p$ form a three-dimensional ``space''.  We take such collections to represent space at a time.

Finally, given space at any time, Galilean spacetime includes a notion of spatial distance between those events, and a notion of angle between arrows relating those events, that satisfy all of the conditions we placed on spatial distance above.  Note, however, that we do not have any notion of spatial distance between points that are not simultaneous---a caveat that will turn out to be important in what follows.

\begin{figure}
\begin{centering}
\includegraphics[width=\columnwidth]{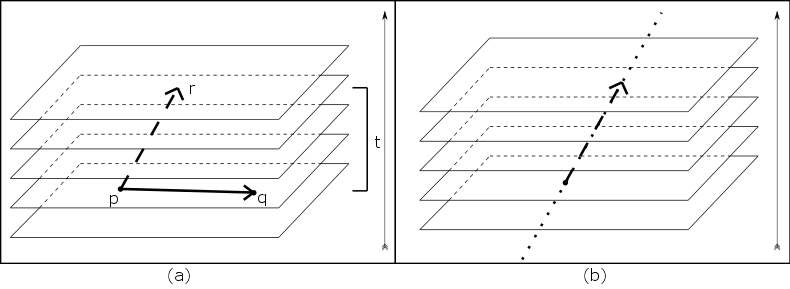}
\end{centering}
\caption{\label{spacetimefig} Galilean spacetime may be thought of as consisting of copies of three-dimensional space stacked on top of one another to form a four-dimensional structure.   (a) Here points $p$ and $q$ are simultaneous, and so the arrow between them has temporal length $0$, but non-zero spatial length; the point $r$ is not simultaneous with $p$ or $q$ and so it has a non-zero temporal distance from both, and its spatial distance to $p$ and $q$ is not defined.  (b) Here we depict the straight line through $p$ determined by a given arrow of unit temporal length.}
\end{figure}

Summing up, we can think of Galilean spacetime as an infinite collection of copies of three-dimensional space, stacked on top of one another to form a four-dimensional structure.\footnote{\label{Galilean} More precisely, Galilean spacetime is a four-dimensional affine space endowed with (1) a non-vanishing linear functional $t$ on its associated vector space; and (2), on each affine subspace by the subspace of vectors to which $t$ assigns $0$, a Eucliean metric $h$.}   (See Fig. \ref{spacetimefig}(a).) Each copy of space has all of the structure we described in the previous section, including the relations of spatial distance and angle.  And between any two slices, we have a notion of duration.  It is in this sense that Galilean spacetime may be taken to represent space and time---i.e., to deserve the name ``spacetime''.

Having described this structure, we can now say the sense in which it is the ``right'' structure for Newtonian physics.  The key is to understand what Galilean spacetime allows one to say about motion.  Consider a particle---i.e., some vanishingly small body that we understand to ``persist'' in the sense of existing over time.  We represent this particle by a collection of events---specifically, by a \emph{curve} through spacetime with the property that it intersects each spatial slice no more than once.\footnote{We intend by ``curve'' a map from (some open subset of) the real numbers into Galilean spacetime that is continuous and at least twice differentiable relative to a topological and differential structure canonically determined by the affine space structure.  For details, see \citet{WeatherallSTG}.}   Such a curve is called a ``trajectory'' or a ``world-line''.  The idea is that the (single) point of intersection between the particle's world-line and any spatial slice represents the location of the particle at that time.  The whole world-line, then, represents the history of the particle over time: it consists in the collection of places, at successive times, that are occupied by the particle.  In other words, it represents the \emph{motion} of the particle through space.

Extended bodies are represented similarly, by ``world-tubes'' that are bounded in each slice of space; the intersection of the world-tube with each slice represents the configuration of the body in space at that instant. In what follows, I will focus on the particle case for simplicity, though much of what is said carries over because one can generally associate a ``center of mass' curve with extended bodies, which characterizes their mass-averaged motion.

We now turn to Newton's laws of motion.  We first remark that, in Galilean spacetime, we have the resources to characterize a special kind of trajectory: namely, a ``straight line through spacetime''.  (See Fig. \ref{spacetimefig}(b).)  Given any point $p$ and any arrow $v$ with temporal length one, we define the \emph{straight line trajectory} through $p$ with \emph{4-velocity} $v$ to consist of all the points one can reach from $p$ by scaling $v$.  (We say 4-velocity because the arrow $v$ relates points in four-dimensional spacetime.)  These trajectories describe motions through spacetime wherein a body moves in some fixed direction at a constant velocity.  Such trajectories play an important role in Newton's theory, encapsulated in his first law of motion.

\begin{quote}\textbf{Newton's First Law}: In the absence of any external force, massive bodies will follow straight line trajectories.\end{quote}

This law establishes a tight connection between two classes of curves: the \emph{mathematically} privileged curves picked out by the structure of Galilean spacetime, and the \emph{physically} privileged curves picked out by the ``default,'' force-free motions of bodies.  These default trajectories are known as \emph{inertial} trajectories.\footnote{See \citet{Earman+Friedman} for a discussion of the status of Newton's first law of motion; see \citet{DiSalleSEP} for a discussion of inertial frames more generally.}

\begin{figure}
\begin{centering}
\includegraphics[width=\columnwidth]{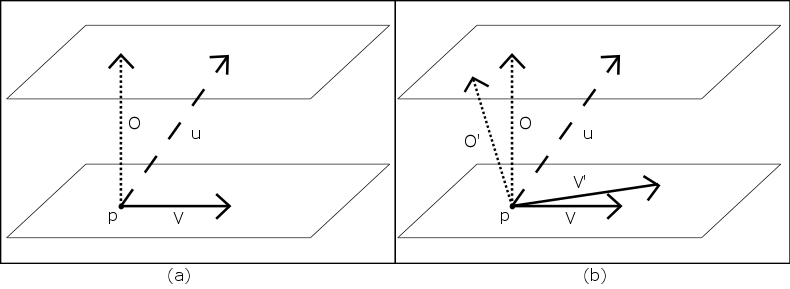}
\end{centering}
\caption{\label{4velocity}  The motion of a body is described by its 4-velocity, which is an arrow $u$ with temporal length $1$. (a) To recover ordinary ``3-velocity'' one introduces an observer $O$ with its own 4-velocity; the (relative) 3-velocity of the body at $p$ is given by $V=u-O$, which is an arrow with temporal length $0$.  (b) Two different observers, $O$ and $O'$, with different states of motion, will determine different 3-velocities, $V$ and $V'$.  It is in this sense that (3-)velocity is \emph{relative} in Galilean spacetime.}
\end{figure}

Before proceeding, let us briefly comment on the role of 4-velocity here.  We usually understand velocity as a quantity with some direction (and magnitude) in \emph{space}; 4-velocity, meanwhile, has direction in \emph{spacetime}.  These are different. To recover velocity as we usually conceive of it---i.e., \emph{3-velocity}---we need to introduce an \emph{observer}, $O$, which is an idealized measuring apparatus, situated somewhere in space and with its own inertial state of motion (represented by some trajectory with 4-velocity, which we also denote $O$).  3-velocity at a time as determined by this observer is given by $u-O$, where $u$ is the 4-velocity of the body at that time. (See Fig. \ref{4velocity}(a).)  This will always be a vector with temporal length $0$ (since $O$ and $u$ both have temporal length one), and so 3-velocity relative to any observer has spatial length, representing the ``speed'' of motion.

The important point about 3-velocity is that it is an essentially \emph{relative} notion: that is, its value at a time depends on the state of motion of the observer.  Different observers, with different states of motion, would attribute different 3-velocities to a body.  (See Fig. \ref{4velocity}(b).)  Indeed, in Galilean spacetime, the notion of an ``absolute'', i.e., non-relative, 3-velocity does not make sense.  But this does not mean that all measures of motion are relative.  In particular, 4-velocity may be characterized independently of any observer.

Thus far, we have focused on straight line trajectories.  But not all trajectories are straight: in general, the motion of a body will change over time.  We capture this by associating an \emph{instantaneous 4-velocity} with a body at each instant.  Given two (necessarily non-simultaneous) points $p$ and $q$ on a body's trajectory, we can define the change in 4-velocity from $p$ to $q$ to be the difference $v(p)-v(q)$, where $v(p)$ is the 4-velocity at $p$ and $v(q)$ is the 4-velocity at $q$; by considering the change in 4-velocity between pairs of points that are ever closer together, and scaling appropriately, we can define the instantaneous rate of change of the particle's 4-velocity, or its \emph{acceleration}, at each point of its worldline.  The acceleration will itself be an arrow at $p$ with temporal length $0$.  Acceleration is important in Newton's second law of motion.

\begin{quote}\textbf{Newton's Second Law}: If a body of mass $m$ is subject to an external force $F$ at a point, then the acceleration $a$ of the body at that point will satisfy $F=ma$.\end{quote}

Thus we see that an impressed force, represented by an arrow $F$, causes a body to deviate from inertial motion, i.e., to accelerate in the direction of the force.  The mass of the body is what determines the magnitude of the acceleration.  It is important to emphasize that acceleration as just defined in Galilean spacetime is an absolute quantity---it is defined without reference to any observer---which is crucial to Newton's second law because force is meant to be an objective, absolute quantity (i.e., not observer dependent), and it would not make sense to have a law that equates a relative quantity with an absolute quantity.

For completeness, we include Newton's third law, although it is not essential to the discussion (and we suppress commentary for space reasons).

\begin{quote}\textbf{Newton's Third Law}: If body 1 exerts a force $F$ on body 2, then body 2 exerts a force $-F$ on body 1.\end{quote}

Finally, although it is not strictly part of Newtonian mechanics, it is worth noting that Newton's laws provide a framework for detailed studies of particular forces.  Newton, for instance, focused on gravitation.

\begin{quote}\textbf{Newton's Law of Universal Gravitation}:  A body of mass $m_1$ located at point $p$ will exert on a body of mass $m_2$ located at (simultaneous) point $q$ a force given by
\[
F=\frac{Gm_1m_2}{||r||^3}r
\]
where $r$ is the arrow from $q$ to $p$, $||r||$ is the spatial length of that arrow, and $G$ is Newton's constant.\end{quote}

It is worth emphasizing that this law invokes temporal length (to define simultaneity) and spatial distance (to define distance between the bodies).  To further relate this force to the acceleration of a body, of course, one also requires the full four dimensional arrow structure of spacetime, as discussed above in connection with Newton's second law.  Thus, we see that all of the structure of Galilean spacetime enters into Newton's laws of motion and his law of universal gravitation.

\section{Newtonian and Leibnizian Spacetimes}

Although the language may be anachronistic, Newton, in the \emph{Principia} and elsewhere,\footnote{See, especially, the Scholium to the Definition, i.e., the Scholium on Space, Time, Place, and Motion \citep[pp. 408-415]{Newton}.} took for granted the structure of Galilean spacetime.  But as \citet{SteinNST} convincingly argues, Newton also believed that spacetime had further structure: namely, what Newton called \emph{absolute space}, which is the structure needed to say whether two places, at different times, are ``the same''.  In other words, Newton believed there was a basic matter of fact about whether any given object remains in one place over time---i.e., at rest---or if it moves.

As we have just seen, this is precisely what Galilean spacetime does \emph{not} provide.  There, one can say whether a body is \emph{accelerating}, i.e., if its velocity is changing over time.  But one cannot say that the body is at rest or moving (at a constant velocity).

To represent this further structure, we need a way of identifying points in space at different times: that is, a way of saying that ``here'' now is the same as ``here'' five minutes ago.  We do so by choosing some arrow $\xi$, of unit temporal length, as special.    This arrow represents Newton's absolute space as follows: two events $p$ and $q$ are at the \emph{same place} (at different times) if and only if the arrow from $p$ to $q$ is some multiple (possibly zero) of $\xi$.  Likewise, a particle is \emph{at rest} (relative to absolute space) if its worldline is the straight line determined by $\xi$.  More generally, a particle with constant velocity (in the sense of Galilean spacetime) has \emph{absolute velocity} $||\xi-u||$, where $u$ is the arrow of temporal length one determining that particle's world-line.  If we take Galilean spacetime, and we add to it this further structure of a privileged arrow $\xi$, we arrive at \emph{Newtonian spacetime}.\footnote{In other words, Newtonian spacetime is Galilean spacetime as described in footnote \ref{Galilean}, with a privileged vector of unit temporal length.}

There are several reasons why someone---even someone who accepted all of Newton's physics---might worry that the world does not exhibit the full structure of Newtonian spacetime.  One particularly influential reason for skepticism, most famously associated with Leibniz's arguments in his correspondence with Clarke \citep[pp. 18f.]{Leibniz}, is that the structure of absolute space, as characterized by $\xi$, has no observable consequences.  In particular, one could take any system of bodies and imagine setting the entire system in motion, at some constant speed (relative to $\xi$) in some fixed direction, and show that, by the lights of Newton's own theory, the relative motions of the bodies would be unchanged.

Newton fully understood this, of course: Corollary 5 to the Laws of Motion \citep[p. 423]{Newton} demonstrates precisely this fact.  But Leibniz took it to have great significance.  He concluded, on its basis, that space and time could not have the structure of absolute space.

But what structure \emph{would} Leibniz attribute to space and time?  It is not entirely clear that a satisfactory or complete answer is to be extracted from Leibniz's writings.  But as \citet{SteinPrehistory} argues, and \citet{Earman1977, Earman1979, Earman1986, EarmanWEST, Earman1989} develops, one might begin with Leibniz's claim, in various places, that \emph{all} motion is relative---i.e., that there are facts only about how the relative configurations of bodies change over time.\footnote{It is not universally accepted that Leibniz was committed to Leibnizian spacetime: see, for instance, \citet{RobertsLeibniz}; see also footnote \ref{EarmanStein}.  For more background on Leibniz's views on physics (and philosophy of physics), see, for instance, \citet{Garber}, \citet{McDonough}, and references therein.}  In other words, it would not make sense to say anything more about the motion of a given body than to describe the rates at which its distance from other bodies is changing over time.

Both Newtonian spacetime and Galilean spacetime provide the resources to describe these changes.  One can consider, for instance, at each time a collection of point-like particles in space, whose relative configuration is represented by the places at which they are located (and the arrows between those places); and one can describe, by word-lines through spacetime, the changes in relative position of these bodies over time.  But one can also say more than this: for instance, in both of these structures, there is a fact about each particle's acceleration---not as a relative matter of the rate of change of the relative velocity of one particle with respect to another, but as an absolute matter.
And this is precisely what someone who thinks all motion is relative would deny.

Thus, to characterize Leibniz's views, we need to \emph{excise} from Galilean spacetime the structure that allowed us to distinguish a special class of motions---namely, the arrows relating non-simultaneous points.  We are left with \emph{Leibnizian spacetime}, which is a collection of events with a notion of temporal distance between pairs of events, such that the collection of all points simultaneous with any given point have the structure of three-dimensional space as described in section \ref{space}, but where we have no arrows between non-simultaneous events.\footnote{More precisely, Leibnizian spacetime is a three dimensional affine bundle over a one dimensional affine space, where each fiber of the bundle is endowed with a Euclidean metric $h$ and there is a one form $t$ on the base space representing temporal distance.}

We saw above that Galilean spacetime---and thus, Newtonian spacetime---provided the resources needed for Newtonian physics.  Can we say the same of Leibnizian spacetime?  In short, no---for reasons Newton himself pointed out.\footnote{Again, see the discussion in the Scholium to the Definitions \citep[pp. 408--415]{Newton}.}  Indeed, Newton offered at least one case in which absolute motion \emph{would} be empirically testable.  Imagine a bucket partially filled with water.  Newton argued that if the bucket is rotating (absolutely) then the surface of the water will appear curved; if the bucket is at rest, then the surface of the water will appear flat. Newton argued that this would be the case, according to his theory, even if there were nothing in the universe but the bucket and the water, and so the behavior of the water in the bucket could not have anything to do with merely \emph{relative} motion.

This thought experiment has been very influential, and Newton himself believed it settled the issue in favor of absolute space---i.e., Newtonian spacetime.  But it is not quite what it seems. The key to understanding the bucket experiment is to realize that any rotating object is accelerating.  In particular, each little bit of the bucket is constantly changing velocity, because it is changing the direction in which it is moving.  Thus Newton's thought experiment is not a way of measuring absolute motion in the sense of measuring absolute velocity.  Instead, it is an experiment to determine (one kind of) absolute \emph{acceleration}.  In other words, the bucket experiment is an argument that we need at least the structure of Galilean spacetime to make sense of Newtonian physics; meanwhile, Leibniz's shift argument against Newtonian spacetime, that there are no empirical tests of absolute \emph{velocity} in Newtonian physics, still stands---as it must, in light of Corollary 5.  Taken together, then, we are pushed to Galilean spacetime as a spacetime structure intermediate between Newtonian and Leibnizian spacetimes.

\section{Maxwell-Huygens and Newton-Cartan Spacetimes}

At the end of the previous section, I observed that in order to accommodate Newton's bucket thought experiment, one needs more structure than Leibnizian spacetime provides.  We went from this remark to the conclusion that one needs the structure of Galilean spacetime (at least) to support Newtonian physics.  But one might worry that this conclusion is too fast: Galilean spacetime makes all acceleration absolute, whereas the bucket thought experiment concerns only a very special kind of accelerated motion: namely, \emph{rotation}.  Perhaps rotation must be absolute to accommodate Newtonian physics, but does it follow that all acceleration is absolute?

This worry is buttressed by the fact that, immediately after Corollary 5 to the Laws of Motion, Newton proves another result: Corollary 6 to the Laws of Motion, which establishes that if a system of bodies is undergoing uniform linear acceleration---i.e., all of the bodies have, in addition to their other motions, some fixed acceleration in a given spatial direction---then their motions relative to one another would be indistinguishable from the case in which that acceleration was absent \citep[p.423ff]{Newton}.  Corollary 6 makes precise a certain sense in which absolute linear acceleration has the same status as absolute velocity.

Very recently, a number of authors have taken up the question of whether Corollary 6 motivated adopting some alternative spacetime structure, intermediate between Leibnizian spacetime and Galilean spacetime, asthe structure presupposed by Newtonian physics.\footnote{Questions about classical spacetime structure in light of Corollary 6 are hardly new: see, for instance, \citet{SteinPrehistory} and \citet{DiSalle}; see also \citet{MalamentCosmology} and \citet{NortonNewton} for an older discussion of the ``relativity of acceleration'' in Newtonian physics in a different context.  But whether Corollary 6 provides an argument against Galilean spacetime has been of particular interest recently \citep{Saunders, KnoxNEP,WeatherallSaunders,WeatherallMaxwell, WallaceEFG, WallaceMPNC, TehRecovery, DewarMaxwell}.}  For instance, Simon \citet{Saunders} argues that a spacetime structure that he calls \emph{Newton-Huygens spacetime} is the proper setting for Newtonian physics.\footnote{Newton-Huygens spacetime was first introduced by \citet{EarmanWEST} under the moniker \emph{Maxwellian spacetime} \citep{EarmanWEST}; \citet{WeatherallSaunders} and others have called it \emph{Maxwell-Huygens} spacetime.}    Newton-Huygens spacetime is like Leibnizian spacetime---i.e., there are notions of spatial and temporal distance, but no arrows between non-simultaneous events---except one has, in addition, a standard for whether a system of bodies is rotating over time.\footnote{In other words, Newton-Huygens spacetime is Leibnizian spacetime endowed with a \emph{standard of rotation} \citep{WeatherallMaxwell}.}  But one does not, in general, have a notion of absolute acceleration.

There are reasons to be cautious about accepting this move, however.  For one, it is more radical than it may at first appear.  In Newtonian physics, forces are not relative: either a body experiences a force or it does not.\footnote{Curiously, Leibniz, too, seems to have taken force to be absolute in this sense, which may raise issues for his views on space and time as reconstructed here.  See \citet{RobertsLeibniz}, \citet{GarberForce}, and \citet{McDonough} for more on Leibniz on force.}  But then, by dint of Newton's second law, acceleration must also not be relative, since it is incoherent to say that an absolute quantity is proportional to a relative quantity.\footnote{As \citet[pp. 19-20]{SteinPrehistory} puts it in a very nice discussion of precisely these issues, absolute acceleration is a \emph{vera causa} in Newtonian physics.}  It follows that to accept Newton-Huygens spacetime, one needs to revise both the conceptual and mathematical foundations of Newtonian physics.\footnote{This is a project that \citet{Saunders} and \citet{DewarMaxwell} have undertaken.}  How significant a revision this amounts to, and whether it can succeed, is a topic of ongoing debate.

A second reason to be cautious is that another response to Corollary 6 is available.  In particular, Eleanor \citet{Knox} has argued that Corollary 6 supports a move to a different theory of gravitation, sometimes known as \emph{geometrized Newtonian gravitation} or \emph{Newton-Cartan theory}, developed by \citet{Cartan1,Cartan2} and \citet{Friedrichs}.\footnote{For background on geometrized Newtonian gravitation, see \citet{Trautman} and \citet[Ch. 4]{MalamentGR}.  For more on the relationship between geometrized Newtonian gravitation and ordinary Newtonian gravitation see, in addition to those resources, \citet{GlymourTE}, \citet{KnoxNEP}, and \citet{WeatherallTheoreticalEquiv}; for discussions of the relationship between geometrized Newtonian gravitation and gravitation in Newton-Huygens spacetime, see \citet{WeatherallSaunders}, \citet{WallaceEFG}, and \citet{DewarMaxwell}.}  Geometrized Newtonian gravitation is a theory with the same empirical consequences as Newtonian gravitation, but set in a spacetime structure importantly different from any of those discussed thus far: it is a theory in which spacetime is \emph{curved} by the distribution of matter in the universe, and where the motion of bodies in spacetime is influenced by that curvature.\footnote{In this, it is like general relativity.  See \citet{Wald} or \citet{MalamentGR} for texbook treatments of general relativity; see also the chapter of this volume on relativistic spacetime.}  The details of geometrized Newtonian gravitation are beyond the scope of this chapter, but one point is worth emphasizing.  In geometrized Newtonian gravitation, while there are (in general) no arrows between non-simultaneous events, there is nonetheless a standard of absolute acceleration.

\section*{Acknowledgments}
This paper is partially based upon work supported by the National Science Foundation under Grant No. 1331126.  I am grateful to David Malament for helpful comments on an earlier draft.

\singlespacing


\begin{thebibliography}{61}
\expandafter\ifx\csname natexlab\endcsname\relax\def\natexlab#1{#1}\fi
\expandafter\ifx\csname url\endcsname\relax
  \def\url#1{\texttt{#1}}\fi
\expandafter\ifx\csname urlprefix\endcsname\relax\def\urlprefix{URL }\fi

\bibitem[{Barrett(2015)}]{BarrettSTS}
Barrett, T., 2015. Spacetime structure. Studies in History and Philosophy of
  Modern Physics 51, 37--43.

\bibitem[{Belot(2000)}]{BelotGM}
Belot, G., 2000. Geometry and motion. British Journal for the Philosophy of
  Science 51~(4), 561--595.

\bibitem[{Cartan(1923)}]{Cartan1}
Cartan, E., 1923. Sur les vari\'et\'es \`a connexion affine, et la th\'eorie de
  la relativit\'e g\'en\'eralis\'ee (premi\`ere partie). Annales scientifiques
  de l'\'Ecole Normale Sup\'erieure 40, 325--412.

\bibitem[{Cartan(1924)}]{Cartan2}
Cartan, E., 1924. Sur les vari\'et\'es \`a connexion affine, et la th\'eorie de
  la relativit\'e g\'en\'eralis\'ee (premi\`ere partie) (suite). Annales
  scientifiques de l'\'Ecole Normale Sup\'erieure 41, 1--25.

\bibitem[{Dewar(2017)}]{DewarMaxwell}
Dewar, N., 2017. Maxwell gravitation, forthcoming in Philosophy of Science.
  Pre-print available at http://philsci-archive.pitt.edu/12470/.

\bibitem[{Dewar and Weatherall(2017)}]{Weatherall+Dewar}
Dewar, N., Weatherall, J.~O., 2017. On gravitational energy in {N}ewtonian
  theories, arXiv:1707.00563 [physics.hist-ph].

\bibitem[{DiSalle(2008)}]{DiSalle}
DiSalle, R., 2008. Understanding Space-Time. Cambridge University Press, New
  York.

\bibitem[{DiSalle(2016)}]{DiSalleSEP}
DiSalle, R., 2016. Space and time: Inertial frames. In: Zalta, E.~N. (Ed.), The
  Stanford Encyclopedia of Philosophy, winter 2016 edition Edition. Available
  at: https://plato.stanford.edu/archives/win2016/entries/spacetime-iframes/.

\bibitem[{Earman(1977)}]{Earman1977}
Earman, J., 1977. Leibnizian space-times and {L}eibnizian algebras. In: Butts,
  R.~E., Hintikka, J. (Eds.), Historical and Philosophical Dimensions of Logic,
  Methodology and Philosophy of Science. Reidel, Dordrecht, pp. 93--112.

\bibitem[{Earman(1979)}]{Earman1979}
Earman, J., 1979. Was {L}eibniz a relationist? Midwest Studies in Philosophy
  4~(1), 263--276.

\bibitem[{Earman(1986)}]{Earman1986}
Earman, J., 1986. Why space is not a substance (at least not to first degree).
  Pacific Philosophical Quarterly 67~(4), 225--244.

\bibitem[{Earman(1989{\natexlab{a}})}]{Earman1989}
Earman, J., 1989{\natexlab{a}}. Leibniz and the absolute vs. relational
  dispute. In: Rescher, N. (Ed.), Leibnizian Inquiries. A Group of Essays.
  University Press of America, Lanham, MD, pp. 9--22.

\bibitem[{Earman(1989{\natexlab{b}})}]{EarmanWEST}
Earman, J., 1989{\natexlab{b}}. World Enough and Space-Time. The MIT Press,
  Cambridge, MA.

\bibitem[{Earman and Friedman(1973)}]{Earman+Friedman}
Earman, J., Friedman, M., 1973. The meaning and status of {N}ewton's law of
  inertia and the nature of gravitational forces. Philosophy of Science 40,
  329.

\bibitem[{Fletcher(2014)}]{FletcherLimit}
Fletcher, S.~C., 2014. On the reduction of general relativity to {N}ewtonian
  gravitation, unpublished manuscript.

\bibitem[{Friedman(1983)}]{Friedman}
Friedman, M., 1983. Foundations of Space-Time Theories: Relativistic Physics
  and Philosophy of Science. Princeton University Press, Princeton, NJ.

\bibitem[{Friedrichs(1927)}]{Friedrichs}
Friedrichs, K.~O., 1927. Eine invariante {F}ormulierung des {N}ewtonschen
  {G}ravitationsgesetzes und der {G}renz\"{u}berganges vom {E}insteinschen zum
  {N}ewtonschen {G}esetz. Mathematische Annalen 98, 566--575.

\bibitem[{Garber(1995)}]{Garber}
Garber, D., 1995. Leibniz: Physics and philosophy. In: Jolley, N. (Ed.), The
  Cambridge Companion to Leibniz. Cambridge University Press, pp. 270--352.

\bibitem[{Garber(2012)}]{GarberForce}
Garber, D., 2012. Leibniz, {N}ewton, and force. In: Janiak, A., Schliesser, E.
  (Eds.), Interpreting {N}ewton. Cambridge University Press, Cambridge, UK, pp.
  33--47.

\bibitem[{Geroch(1981)}]{GerochAB}
Geroch, R., 1981. General Relativity from A to B. University of Chicago Press,
  Chicago, IL.

\bibitem[{Glymour(1980)}]{GlymourTE}
Glymour, C., 1980. Theory and Evidence. Princeton University Press, Princeton,
  NJ.

\bibitem[{Knox(2011)}]{Knox}
Knox, E., 2011. {N}ewton-{C}artan theory and teleparallel gravity: The force of
  a formulation. Studies in History and Philosophy of Modern Physics 42~(4),
  264--275.

\bibitem[{Knox(2014)}]{KnoxNEP}
Knox, E., 2014. {N}ewtonian spacetime structure in light of the equivalence
  principle. The British Journal for the Philosophy of Science 65~(4),
  863--888.

\bibitem[{Leibniz and Clarke(2000)}]{Leibniz}
Leibniz, G.~W., Clarke, S., 2000. Correspondence. Hackett Publishing Co.,
  Indianapolis, IN, trans. and ed. by Roger Ariew.

\bibitem[{Malament(1986{\natexlab{a}})}]{MalamentLimit1}
Malament, D., 1986{\natexlab{a}}. Gravity and spatial geometry. In: Marcus,
  R.~B., Dorn, G., Weingartner, P. (Eds.), Logic, Methodology and Philosophy of
  Science. Vol. VII. Elsevier Science Publishers, New York, pp. 405--411.

\bibitem[{Malament(1986{\natexlab{b}})}]{MalamentLimit2}
Malament, D., 1986{\natexlab{b}}. {N}ewtonian gravity, limits, and the geometry
  of space. In: Colodny, R. (Ed.), From Quarks to Quasars. University of
  Pittsburgh Press, Pittsburgh, pp. 181--201.

\bibitem[{Malament(1995)}]{MalamentCosmology}
Malament, D., 1995. Is {N}ewtonian cosmology really inconsistent? Philosophy of
  Science 62~(4), 489--510.

\bibitem[{Malament(2009)}]{MalamentGST}
Malament, D.~B., 2009. Notes on geometry and spacetime, unpublished lecture
  notes. Available at:
  http://www.socsci.uci.edu/~dmalamen/courses/geometryspacetimedocs/GST.pdf.

\bibitem[{Malament(2012)}]{MalamentGR}
Malament, D.~B., 2012. Topics in the Foundations of General Relativity and
  {N}ewtonian Gravitation Theory. University of Chicago Press, Chicago.

\bibitem[{Maudlin(2012)}]{Maudlin}
Maudlin, T., 2012. Philosophy of physics: Space and time. Princeton University
  Press, Princeton, NJ.

\bibitem[{McDonough(2014)}]{McDonough}
McDonough, J.~K., 2014. Leibniz's philosophy of physics. In: Zalta, E.~N.
  (Ed.), The Stanford Encyclopedia of Philosophy, spring 2014 ed. Edition.
  Available at:
  https://plato.stanford.edu/archives/spr2014/entries/leibniz-physics/.

\bibitem[{Minkowski(2013 [1911])}]{Minkowski}
Minkowski, H., 2013 [1911]. Space and time. In: Lewertoff, F., Petkov, V.
  (Eds.), Space and Time: Minkowski's Papers on Relativity. Minkowski Institute
  Press, Montreal, pp. 111--125.

\bibitem[{Newton(1999 [1687 / 1713 / 1726])}]{Newton}
Newton, I., 1999 [1687 / 1713 / 1726]. The Principia: Mathematical Principles
  of Natural Philosophy. University of California Press, Berkeley, CA, edited
  and trans. by I. Bernard Cohen and Anne Whitman.

\bibitem[{Norton(1995)}]{NortonNewton}
Norton, J.~D., 1995. The force of {N}ewtonian cosmology: Acceleration is
  relative. Philosophy of Science 62~(4), 511--522.

\bibitem[{Pooley(2013)}]{PooleyOxford}
Pooley, O., 2013. Substantivalist and relationalist approaches to spacetime.
  In: Batterman, R. (Ed.), The Oxford Handbook of Philosophy of Physics. Oxford
  University Press, Oxford, UK, pp. 522--586.

\bibitem[{Roberts(2003)}]{RobertsLeibniz}
Roberts, J.~T., 2003. Leibniz on force and absolute motion. Philosophy of
  Science 70~(3), 553--573.

\bibitem[{Rosenstock et~al.(2015)Rosenstock, Barrett, and
  Weatherall}]{Rosenstock+etal}
Rosenstock, S., Barrett, T., Weatherall, J.~O., 2015. On einstein algebras and
  relativistic spacetimes. Studies in History and Philosophy of Modern Physics
  52B, 309--316.

\bibitem[{Rynasiewicz(1992)}]{Rynasiewicz1992}
Rynasiewicz, R., 1992. Rings, holes and substantivalism: On the program of
  {L}eibniz algebras. Philosophy of Science 59~(4), 572--589.

\bibitem[{Saunders(2013)}]{Saunders}
Saunders, S., 2013. Rethinking {N}ewton's \emph{Principia}. Philosophy of
  Science 80~(1), 22--48.

\bibitem[{Stan(2016)}]{Stan}
Stan, M., 2016. Huygens on inertial structure and relativity. Philosophy of
  Science 83~(2), 277--298.

\bibitem[{Stein(1967)}]{SteinNST}
Stein, H., 1967. {N}ewtonian space-time. The Texas Quarterly 10, 174--200.

\bibitem[{Stein(1977)}]{SteinPrehistory}
Stein, H., 1977. Some philosophical prehistory of general relativity. In:
  Earman, J., Glymour, C., Stachel, J. (Eds.), Foundations of Space-Time
  Theories. University of Minnesota Press, Minneapolis, MN, pp. 3--49.

\bibitem[{Teh(2017)}]{TehRecovery}
Teh, N., 2017. Recovering recovery: On the relationship between gauge symmetry
  and trautman recovery, forthcoming in Philosophy of Science.

\bibitem[{Trautman(1965)}]{Trautman}
Trautman, A., 1965. Foundations and current problem of general relativity. In:
  Deser, S., Ford, K.~W. (Eds.), Lectures on General Relativity. Prentice-Hall,
  Englewood Cliffs, NJ, pp. 1--248.

\bibitem[{Wald(1984)}]{Wald}
Wald, R., 1984. General Relativity. University of Chicago Press, Chicago.

\bibitem[{Wallace(2016)}]{WallaceEFG}
Wallace, D., 2016. Fundamental and emergent geometry in {N}ewtonian physics,
  http://philsci-archive.pitt.edu/12497/.

\bibitem[{Wallace(2017)}]{WallaceMPNC}
Wallace, D., 2017. More problems for {N}ewtonian cosmology. Studies in History
  and Philosophy of Modern Physics 57, 35--40.

\bibitem[{Weatherall(2011{\natexlab{a}})}]{WeatherallExplanation}
Weatherall, J.~O., 2011{\natexlab{a}}. On (some) explanations in physics.
  Philosophy of Science 78~(3), 421--447.

\bibitem[{Weatherall(2011{\natexlab{b}})}]{WeatherallSGP}
Weatherall, J.~O., 2011{\natexlab{b}}. On the status of the geodesic principle
  in {N}ewtonian and relativistic physics. Studies in the History and
  Philosophy of Modern Physics 42~(4), 276--281.

\bibitem[{Weatherall(2014)}]{WeatherallSingularity}
Weatherall, J.~O., 2014. What is a singularity in geometrized {N}ewtonian
  gravitation? Philosophy of Science 81~(5), 1077--1089.

\bibitem[{Weatherall(2015{\natexlab{a}})}]{WeatherallTheoreticalEquiv}
Weatherall, J.~O., 2015{\natexlab{a}}. Are {N}ewtonian gravitation and
  geometrized {N}ewtonian gravitation theoretically equivalent?
  ErkenntnisPublished online. doi:10.1007/s10670-015-9783-5.

\bibitem[{Weatherall(2015{\natexlab{b}})}]{WeatherallHoleArg}
Weatherall, J.~O., 2015{\natexlab{b}}. Regarding the `{H}ole {A}rgument'. The
  British Journal for Philosophy of Science. Forthcoming. arXiv:1412.0303
  [physics.hist-ph].

\bibitem[{Weatherall(2016{\natexlab{a}})}]{WeatherallSaunders}
Weatherall, J.~O., 2016{\natexlab{a}}. Maxwell-{H}uygens, {N}ewton-{C}artan,
  and {S}aunders-{K}nox spacetimes. Philosophy of Science 83~(1), 82--92.

\bibitem[{Weatherall(2016{\natexlab{b}})}]{WeatherallSTG}
Weatherall, J.~O., 2016{\natexlab{b}}. Space, time, and geometry from {N}ewton
  to {E}instein, feat. {M}axwell, lecture notes from the 2016 MCMP summer
  school in mathematical philosophy; available on request.

\bibitem[{Weatherall(2016{\natexlab{c}})}]{WeatherallVoid}
Weatherall, J.~O., 2016{\natexlab{c}}. Void: The Strange Physics of Nothing.
  Yale University Press, New Haven, CT.

\bibitem[{Weatherall(2017{\natexlab{a}})}]{WeatherallMaxwell}
Weatherall, J.~O., 2017{\natexlab{a}}. A brief comment on
  {M}axwell(/{N}ewton){[-Huygens]} spacetime, arXiv:1707.02393
  [physics.hist-ph].

\bibitem[{Weatherall(2017{\natexlab{b}})}]{WeatherallCategories}
Weatherall, J.~O., 2017{\natexlab{b}}. Categories and the foundations of
  classical field theories. In: Landry, E. (Ed.), Categories for the Working
  Philosopher. Oxford University Press, Oxford, UK, arXiv:1505.07084
  [physics.hist-ph].

\bibitem[{Weatherall(2017{\natexlab{c}})}]{WeatherallConservation}
Weatherall, J.~O., 2017{\natexlab{c}}. Conservation, inertia, and spacetime
  geometry, forthcoming in {Studies in History and Philosophy of Modern
  Physics}.

\bibitem[{Weatherall(2017{\natexlab{d}})}]{WeatherallPuzzleBall}
Weatherall, J.~O., 2017{\natexlab{d}}. Inertial motion, explanation, and the
  foundations of classical space-time theories. In: Lehmkuhl, D., Schiemann,
  G., Scholz, E. (Eds.), Towards a Theory of Spacetime Theories. Birkh\"auser,
  Boston, MA, pp. 13--42, arXiv:1206.2980 [physics.hist-ph].

\bibitem[{Weatherall(2017{\natexlab{e}})}]{WeatherallEarmanStein}
Weatherall, J.~O., 2017{\natexlab{e}}. Some philosophical prehistory of the
  earman-norton hole argument, unpublished manuscript.

\bibitem[{Weyl(1952 [1918])}]{Weyl}
Weyl, H., 1952 [1918]. Space Time Matter. Dover Publications, Mineola, NY.

\end{thebibliography}
\end{document}